# High Speed Contact-Resonance Tracking using Brownian motion

*J. Bemis and R. Proksch **

Asylum Research, Oxford Instruments, Santa Barbara, CA, USA

*Author to whom correspondence should be addressed: roger.proksch@oxinst.com

**Abstract**

Contact-resonance atomic force microscopy (CR-AFM) provides nanoscale maps of contact stiffness, dissipation, and electromechanical response, but conventional piezoacoustic excitation can obscure the cantilever resonance with actuator and sample-holder modes. Pure Brownian excitation avoids this transfer-function background, yet its small amplitude generally requires averaging that is incompatible with routine imaging. We introduce interferometric dual-AC resonance tracking (iDART), which combines quadrature-phase differential interferometry with two-frequency resonance tracking. By positioning the interferometric spot near the displacement maximum of the first contact-resonance mode, the detector noise floor is reduced below the off-resonance thermal displacement of the cantilever. Brownian spectra identify the contact mode and define the tracking frequencies, while active electrical excitation provides the signal-to-noise ratio required for pixel-resolved imaging. Simultaneous low-frequency and resonant measurements show that both electrical and photothermal drive increases contrast in the amplitude and phase channels without appreciably shifting the contact resonance. These results establish a practical route to resonance-enhanced, drive free nanomechanical imaging at conventional AFM scan rates.

Contact-resonance atomic force microscopy (CR-AFM) infers local mechanical and electromechanical properties from the resonance frequency, linewidth, amplitude, and phase of a tip-coupled cantilever mode, an approach with roots in early atomic force acoustic microscopy[1, 2] and ultrasonic force microscopy. [3], [4] Operating at or near the contact resonance is equally central to piezoresponse force microscopy (PFM), [5] where resonance enhancement amplifies picometer-scale electromechanical displacements but also entangles them with the position-dependent dynamics of the cantilever–sample contact. Because the contact-resonance frequency shifts as the local contact stiffness varies, resonant imaging requires the drive frequency to follow the resonance, which is accomplished by feedback schemes such as dual-AC resonance tracking (DART) [6,7,8] or by band-excitation methods. [9] The quantitative accuracy of all of these approaches depends on separating the intrinsic cantilever–contact response from the excitation transfer function. Piezoacoustic actuation drives the sample, holder, and cantilever base and can therefore generate a forest of parasitic peaks, particularly in liquid[10] or on mechanically complex sample assemblies. [11], Direct photothermal[12] or magnetic excitation produces cleaner spectra but requires additional hardware and may perturb the cantilever or the contact. Brownian motion provides an actuator-independent spectrum and is therefore attractive for identifying and calibrating the contact resonance;[13] however, the thermal displacement is usually too small for resonance tracking at millisecond pixel dwell times.

Here we use the Brownian spectrum to identify the contact mode and the optimum detection position, and then use a controlled active drive for rapid imaging. The resulting method, iDART, [14] combines the displacement sensitivity of quadrature-phase differential interferometry (QPDI) [15],[16] with dual-AC resonance tracking. Unlike optical beam-deflection (OBD) detection, which measures the local slope of the cantilever, QPDI measures its local displacement. [15] This distinction is important in contact because the tip boundary condition can suppress the slope signal near the tip while leaving a displacement antinode on the cantilever body.[16] Positioning the QPDI spot at that antinode yields a measurable contact-resonance peak with a detector noise floor below the off-resonance Brownian motion (see Fig. 1 and S1 of the supplementary material). iDART was recently shown to provide a tenfold or greater signal-to-noise improvement over state-of-the-art PFM at low drive bias;[14] here we show that this sensitivity advantage extends to the fundamental limit set by the thermal Brownian motion of the cantilever itself.

Measurements were performed with a Vero interferometric AFM (Asylum Research, Oxford Instruments) using controller software version 21.12.79 and modified signal routing that supplied the QPDI displacement channels to the standard DART and vertical-feedback inputs. We used an Adama 2.8 diamond-coated cantilever with a spring constant of 3.15 N/m and a free resonance frequency of ≈72 kHz.

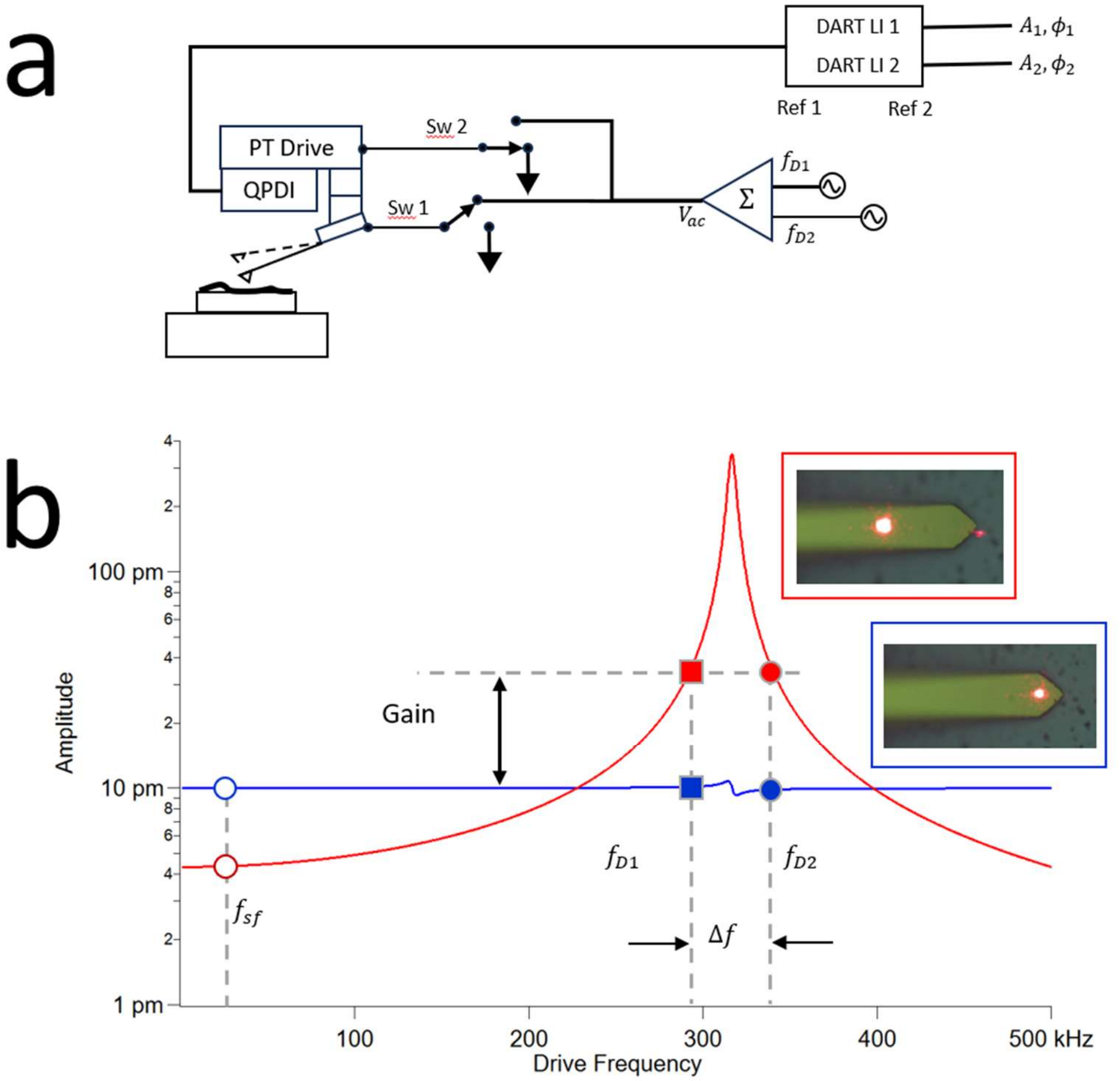

*FIG. 1. (a) Experimental configuration for simultaneous low-frequency interferometric single-frequency (iSF) and interferometric dual-AC resonance-tracking (iDART) measurements. The response is measured at a subresonant frequency fSF and at two frequencies, fD1 and fD2, placed on opposite sides of the contact resonance and separated by Δf. (b) Schematic response for detection near the tip and near the displacement maximum on the cantilever body. The off-resonance responses are comparable, whereas the body-position response is strongly amplified near contact resonance. Insets show the corresponding laser-spot positions.*

Figure 1 summarizes the measurement principle. As shown in Fig. 1(a), two drive tones at frequencies $f_{D1}$ and $f_{D2}$, separated by $\Delta f$ and placed on opposite sides of the contact resonance, are summed and routed by switches either to the tip as an electrical bias $V_{ac}$ or to the photothermal drive. Two lock-in amplifiers demodulate the QPDI displacement signal at these frequencies, returning the amplitude and phase pairs ($A_1$, $\phi_1$) and ($A_2$, $\phi_2$) used by the DART feedback loop to keep the drive pair centered on the resonance. [6,7] A third, subresonant channel at $f_{SF}$ provides a nonresonant interferometric single-frequency (iSF) reference. Recording all three frequencies during the same scan avoids registration errors and isolates the gain produced by resonance enhancement, illustrated schematically in Fig. 1(b). The spot-position dependence, shown in the insets of Fig.

1(b), is central: a position that is nearly insensitive to the contact resonance in the conventional slope signal can coincide with a displacement maximum in QPDI.

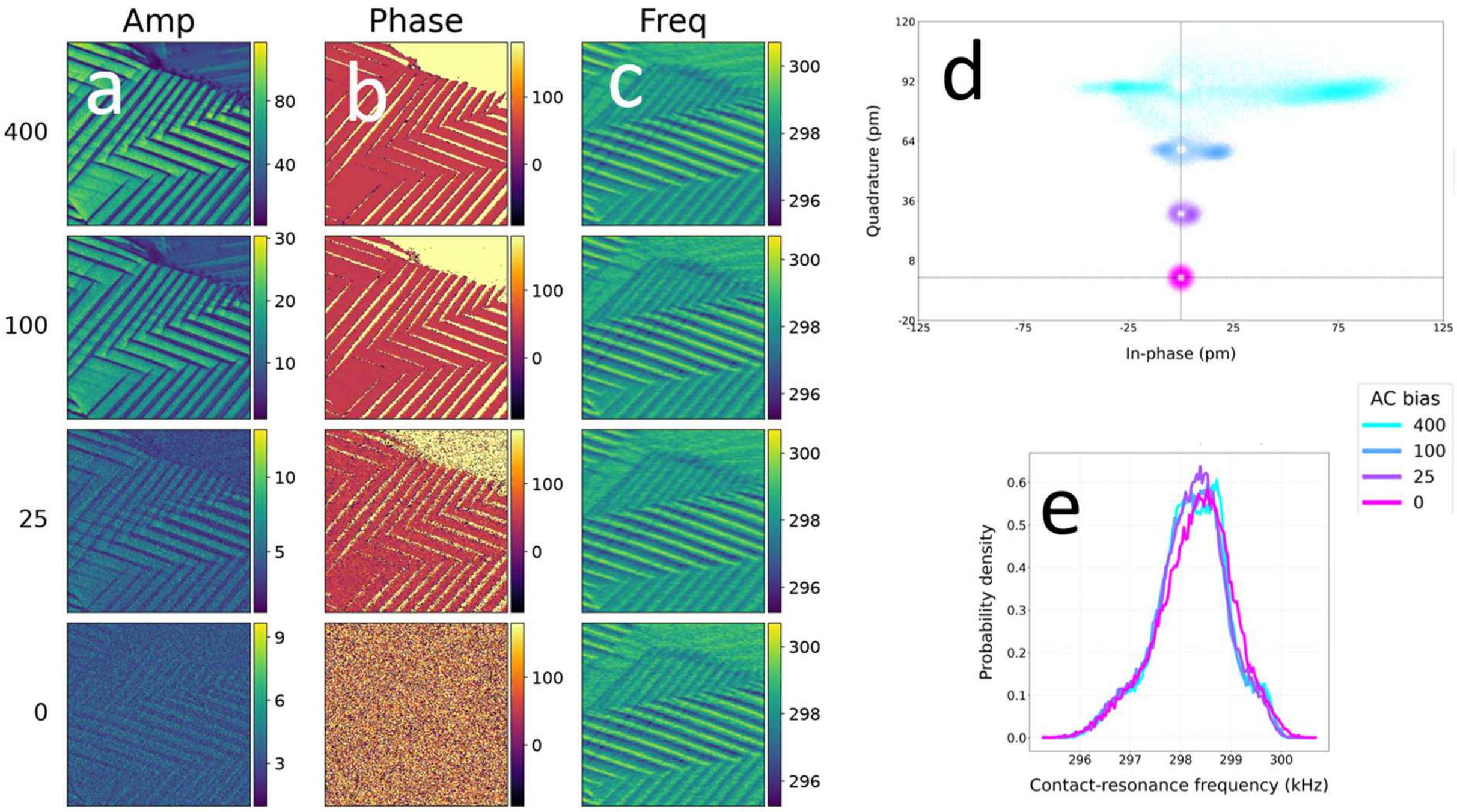

***FIG. 2.*** *iDART PFM measurements of a PZT sample acquired at AC bias amplitudes of 400, 100, 25, and 0 mV. Spatial maps show the PFM amplitude (a), phase (b), and contact-resonance frequency (c). The corresponding complex PFM response is plotted in quadrature space for each AC bias (d). The point clouds are offset vertically along the quadrature axis for clarity. Probability-density distributions of the contact-resonance frequency show that the frequency contrast and distribution remain largely unchanged as the AC bias is reduced to zero (e).*

Figure 2 compares the iDART response acquired at four AC bias amplitudes, $V_{\mathrm{ac}} = 400$, 100, 25, and 0 mV. The piezoresponse amplitude maps in Fig. 2(a) show strong domain contrast at 400 mV, with the alternating domain structure clearly resolved across the field of view. As $V_{\mathrm{ac}}$ is reduced to 100 and 25 mV, the measured amplitude decreases approximately in proportion to the excitation and the domain contrast becomes progressively less distinct relative to the background fluctuations. At $V_{\mathrm{ac}} = 0$, the residual amplitude is dominated by thermally driven cantilever motion and instrumental noise, although faint spatial structure may remain because the plotted amplitude is a positive-definite quantity and therefore does not average to zero.

The phase maps in Fig. 2(b) exhibit the expected approximately 180°contrast between oppositely polarized domains at the larger drive amplitudes. This phase reversal is consistent with a piezoelectric displacement whose sign changes with the polarization direction. At $V_{\mathrm{ac}} = 100$ mV, the domain pattern remains readily identifiable, whereas at $V_{\mathrm{ac}} = 25$ mV, the phase becomes increasingly dispersed as the coherent electromechanical response approaches the stochastic Brownian displacement of the cantilever. At zero applied bias, no phase-coherent piezoresponse is expected, and the phase image becomes effectively random. The loss of phase contrast therefore reflects the decreasing signal-to-noise ratio of the driven piezoelectric response rather than a change in the underlying domain configuration.

In contrast, the iDART-tracked contact-resonance-frequency maps in Fig. 2(c) remain spatially reproducible throughout the entire bias series. The same frequency features are observed at $V_{\mathrm{ac}} = 400$, 100, 25, and 0 mV, even after the amplitude and phase contrast have become strongly degraded. Because the contact-resonance frequency is determined by the local dynamic boundary conditions of the tip–sample contact, these maps primarily reflect variations in contact stiffness, local elastic properties, contact geometry, adhesion, and preload. Their persistence at zero applied AC bias shows that the frequency contrast does not require a coherent electrically driven response and can instead be recovered from the thermally excited cantilever motion. [13]

The quadrature representation in Fig. 2(d) further illustrates the transition from coherent piezoelectric response to Brownian-dominated motion. At 400 mV, the response forms two well-separated populations corresponding to oppositely polarized domains. These populations are displaced mainly along the quadrature direction and are consistent with responses that differ by approximately $180°$in phase. As the AC bias is reduced, the separation between the populations decreases and the distributions contract toward the origin. At 25 mV, the polarization-dependent populations are only weakly separated, while at 0 mV, the response collapses near the origin because there is no coherent piezoelectric excitation. The quadrature plot therefore makes clear that the electrically driven vector response vanishes with decreasing bias even though the contact resonance remains measurable.

The contact-resonance-frequency probability-density distributions in Fig. 2(e) provide a statistical comparison of the frequency measurements. The distributions measured at all four AC bias values overlap closely, with similar peak positions, widths, and overall shapes. No systematic frequency shift is observed as $V_{\mathrm{ac}}$ is reduced from 400 mV to zero. This agreement confirms that the applied electrical excitation does not measurably perturb the local contact resonance over the range investigated. It also demonstrates that iDART can determine the contact-resonance frequency from Brownian motion alone, even when the amplitude and phase are incoherent.

Figure 2 separates two distinct components of the measurement. The amplitude, phase, and quadrature response depend directly on the coherent piezoelectric excitation and therefore vanish as $V_{\mathrm{ac}} \to 0$. The contact-resonance frequency, by comparison, is encoded in both the driven and thermally excited cantilever motion and remains accessible without an applied AC bias. This provides a route to measuring spatial variations in contact stiffness without requiring a large electrical or mechanical excitation that could otherwise introduce electrostatic forces, Joule heating, charge injection, ionic motion, or tip-induced modification of the sample.

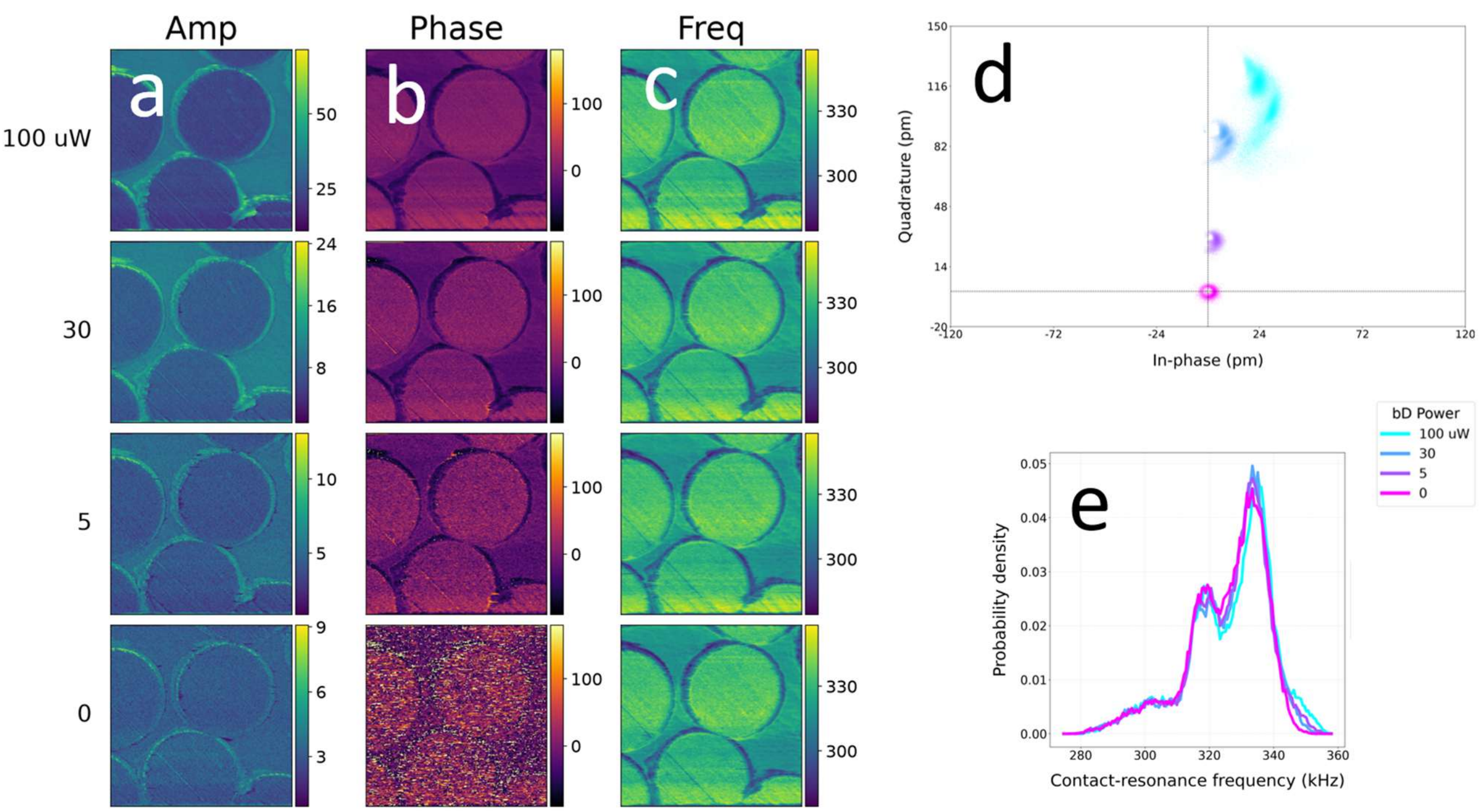

*FIG. 3. Photothermally driven contact-resonance measurements of carbon fibers embedded in an epoxy matrix, acquired at photothermal modulation powers of 100, 30, 5, and* $0\ \mu W$. *Spatial maps show the response amplitude (a), phase (b), and contact-resonance frequency (c). The corresponding complex response is plotted in quadrature space for each photothermal power (d). Probability-density distributions of the contact-*

*resonance frequency (e) show that the principal frequency populations and spatial frequency contrast remain largely unchanged as the photothermal excitation is reduced, including the zero-power control.*

Figure 3 shows a similar crossover from coherent to stochastic response for a sample of carbon fibers embedded in an epoxy matrix, imaged using photothermally excited contact resonance. Photothermal actuation is generated by modulating the output of a laser diode focused near the base of the cantilever. The photothermally driven amplitude maps in Fig. 3(a) show a strong, spatially coherent response at $100\ \mu$W. The boundaries and interiors of the circular features are readily distinguished, indicating that the photothermal excitation efficiently drives the contact-resonance mode. As the modulation power is reduced to 30 and $5\ \mu$W, the measured amplitude decreases and the contrast becomes progressively weaker relative to the thermally driven background. At zero modulation power, the remaining amplitude is largely determined by Brownian cantilever motion, detector noise, and any residual response associated with static illumination or instrumental background.

The phase maps in Fig. 3(b) show a corresponding loss of coherence as the photothermal drive is reduced. At $100\ \mu$W, the phase is spatially well defined, with distinct phase contrast associated with the circular structures and their boundaries. The phase remains interpretable at $30\ \mu$W, but becomes more dispersed at $5\ \mu$Was the coherent response approaches the level of the stochastic thermal motion. In the zero-power control, no phase-locked photothermal response is expected, and the phase becomes effectively random. The disappearance of coherent phase contrast therefore confirms that the externally driven oscillation has been removed.

In contrast, the contact-resonance-frequency maps in Fig. 3(c) retain the same principal spatial features throughout the entire photothermal-power series. The circular regions, their boundaries, and the surrounding matrix remain distinguishable at 100, 30, 5, and $0\ \mu$W, even after the amplitude and phase responses have become weak or incoherent. This persistence shows that the frequency contrast is not generated by the modulated photothermal excitation itself. Instead, it reflects the local dynamic properties of the cantilever–sample contact, including contact stiffness, local elastic response, contact geometry, adhesion, and preload.

The fact that these features remain measurable at zero modulation demonstrates that the contact resonance can be recovered from the thermally excited cantilever motion.

The quadrature representation in Fig. 3(d) further illustrates the transition from a coherent photothermally driven response to Brownian-dominated motion. At $100\ \mu$W, the response occupies a broad, well-separated region in the complex plane, indicating a large coherent oscillation with a defined phase. As the photothermal power is reduced, the response distributions contract toward the origin, consistent with the decreasing driven amplitude. At the lowest powers, the distributions are concentrated near the origin because the coherent oscillation has become small compared with the stochastic thermal motion. The quadrature plot therefore confirms that the driven vector response vanishes with decreasing photothermal power, even though the mechanical resonance remains detectable.

The contact-resonance-frequency probability-density distributions in Fig. 3(e) provide a statistical comparison of the frequency response at each photothermal power. The principal peaks and shoulders remain present at all drive levels, showing that the same mechanically distinct regions are sampled throughout the series. Small changes in the peak positions and relative weights may arise from static laser heating, which can alter the cantilever modulus, normal load, tip–sample contact area, sample modulus, or interfacial conditions. In particular, when the modulation amplitude is reduced to zero while the diode remains DC biased, the stationary laser spot can still produce a finite temperature offset near the cantilever base. Removing the DC bias as well eliminates this residual static heating, allowing the contribution of laser-induced temperature bias to be separated from the effect of dynamic actuation.

As in Fig. 2, the resonance frequency itself remains encoded in the Brownian motion of the cantilever and can still be determined after both the modulated drive and the static optical heating are removed. These results demonstrate that contact-stiffness contrast remains accessible in the absence of external excitation, enabling minimally perturbative or fully passive contact-resonance measurements.

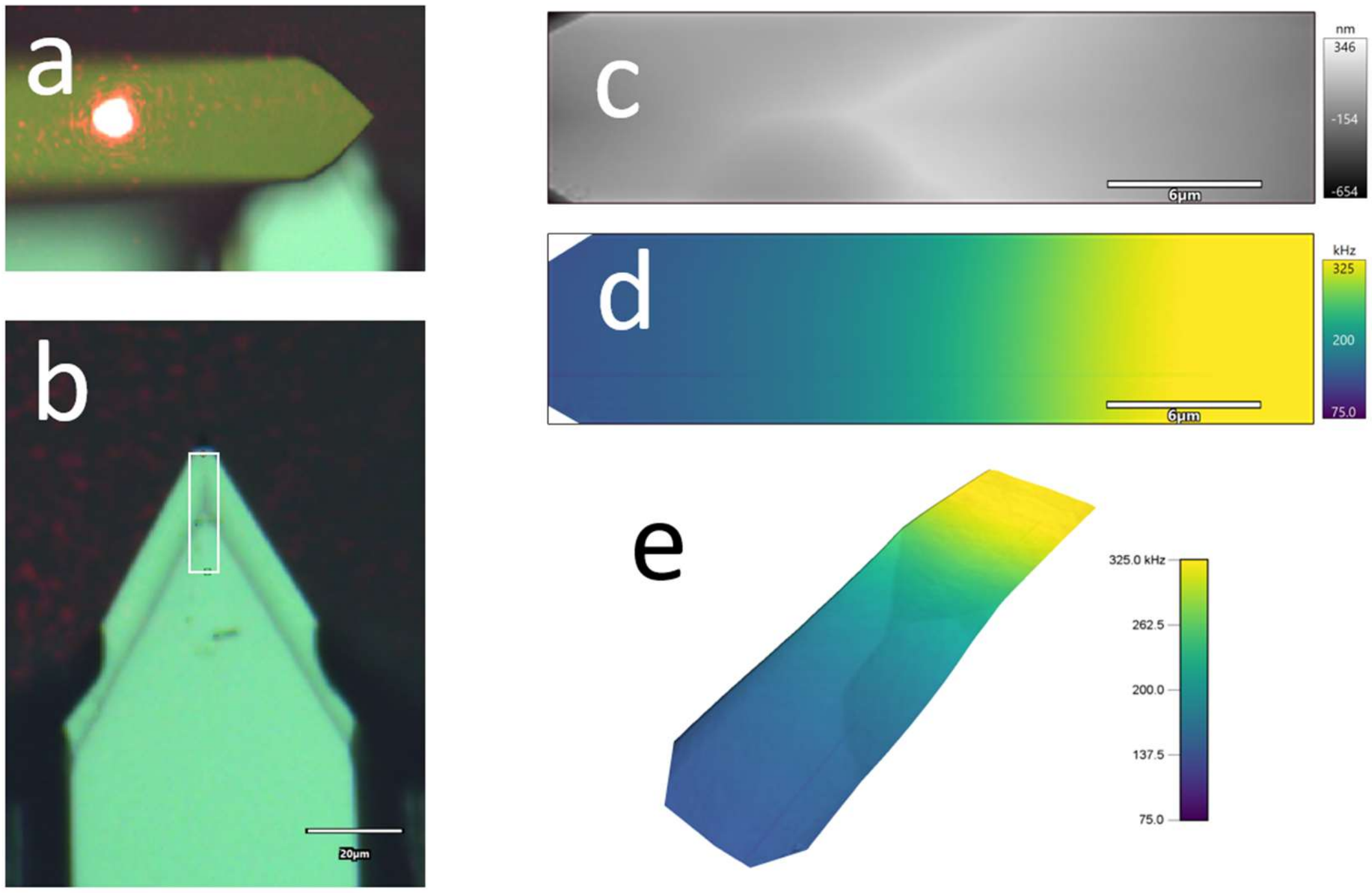

***FIG. 4.*** *Experimental mapping of the position-dependent contact resonance across a triangular cantilever. (a) Optical image showing the QPDI measurement spot positioned near the tip of the upper cantilever while it is brought into contact with the sample cantilever shown below. (b) Top-view optical image of the sample cantilever; the white rectangle marks the region scanned. (c) Measured topography of the selected region. (d) Corresponding contact-resonance-frequency map, spanning approximately* $75 to\ 325\ kHz$. *(e) Three-dimensional rendering of the contact-resonance frequency overlaid on the measured topography, highlighting the systematic increase in resonance frequency along the cantilever.*

The spatially resolved measurements in Fig. 4 provide a direct experimental demonstration of the large dynamic range accessible with iDART. Figure 4(a) shows the QPDI spot positioned near the tip of the measurement cantilever as it is brought into contact with a second, sample cantilever, and the white rectangle in the top-view optical image of Fig. 4(b) marks the region scanned. Across a scan spanning only a few tens of micrometers, the measured contact-resonance frequency changes from approximately $75 \text{to } 325 \text{ kHz}$, corresponding to more than a fourfold variation in frequency. This wide range arises because the local mechanical response depends strongly on position along the triangular cantilever. Near the compliant end, the sample cantilever contributes substantial local deformation and the measured resonance is relatively low. Closer to the wider and more strongly supported region, the local stiffness increases and the contact resonance shifts to much higher frequency.

The topography in Fig. 4(c) shows that the scanned region is not geometrically uniform, while the frequency map in Fig. 4(d) reveals a smooth and much larger mechanical gradient across the same area. The three-dimensional representation in Fig. 4(e) makes clear that the resonance-frequency variation is systematic rather than random and is correlated with position along the cantilever. The ability to track the resonance continuously over this range demonstrates that iDART is not limited to small perturbations around a single nominal contact resonance.

These data also emphasize that the measured response is determined by the combined cantilever–contact system. Changes in local width, thickness, support conditions, and mode shape alter the effective dynamic stiffness and therefore the resonance frequency. For a short triangular cantilever with nonuniform geometry, this behavior cannot be captured adequately by a simple uniform Euler–Bernoulli beam approximation. Nevertheless, iDART remains able to follow the evolving resonance over the full range, illustrating its utility for samples with large spatial variations in stiffness and strongly position-dependent mechanical boundary conditions.

In summary, iDART uses interferometric displacement detection at or near a contact-mode antinode to recover resonance information that is inaccessible to conventional tip-position OBD detection. Brownian spectra provide a clean mode-identification and frequency-selection step, while active electrical excitation supplies the amplitude required for rapid pixel-by-pixel tracking. The measured frequency is stable over the electrical-drive range tested, while static heating associated with the photothermal drive can introduce small systematic frequency offsets, an experimentally useful distinction between signal enhancement and contact modification. These results directly support the sensitivity improvement reported in the original iDART work,[14] and extend it to its fundamental limit: with the detector noise floor below the thermal displacement of the cantilever, the contact resonance remains trackable with no external drive at all.

**ACKNOWLEDGMENTS**

**AUTHOR DECLARATIONS**

**Conflict of Interest**

The authors are employees of Asylum Research, Oxford Instruments, the manufacturer of the atomic force microscope used in this work.

**Author Contributions**

## DATA AVAILABILITY

The data that support the findings of this study are available from the corresponding author upon reasonable request.

## SUPPLEMENTARY MATERIAL

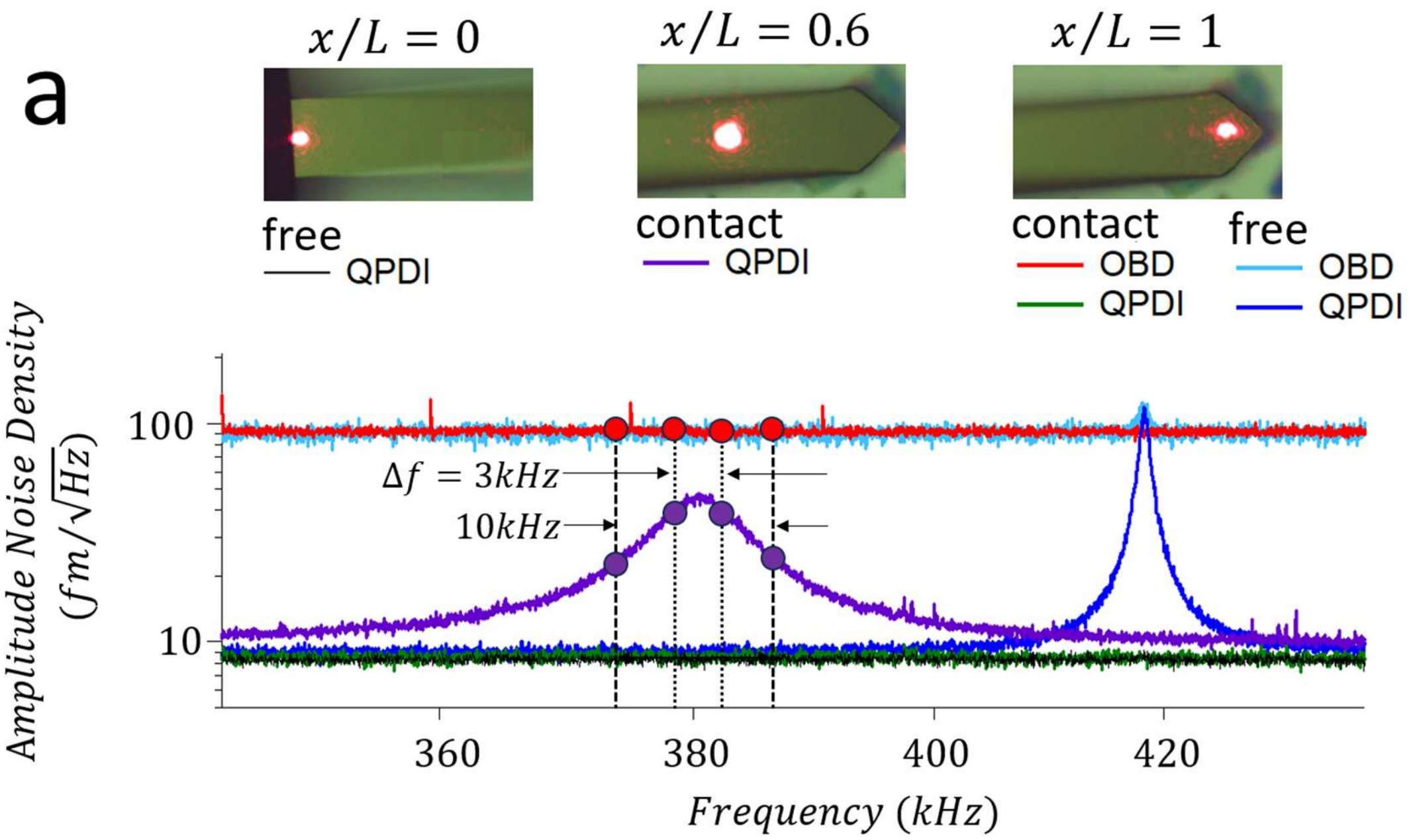

*FIG. S1. Brownian amplitude-noise density for free and contact conditions measured by optical beam deflection (OBD) and QPDI at representative positions along the cantilever. Near the tip, at x/L ≈ 1, both methods resolve the second free resonance mode near 418 kHz (light blue for OBD and dark blue for QPDI), though the signal-to-noise ratio of the QPDI measurement is much larger than that of the OBD measurement. In contact, the signals measured near the tip by both QPDI and OBD (green and red, respectively) are dominated by detector noise. The QPDI signal at x/L ≈ 0.6 (purple) resolves a broad contact resonance near 380 kHz. Symbols on the purple curve represent DART drive pairs separated by Δf = 3 and 10 kHz.*

The undriven spectra in Fig. S1 separate detector noise from Brownian-driven cantilever dynamics. The noise floor of the QPDI detector is approximately $\approx 8\ fm/\sqrt{Hz}$, compared to the $\approx 100\ fm/\sqrt{Hz}$ for OBD detection. As a consequence, when the tip is in contact, OBD with the spot positioned at x/L ≈ 1 does not reveal the Brownian-driven resonance peak. In contrast, the QPDI measurement at x/L ≈ 0.6 clearly reveals the Brownian-driven resonance near 380 kHz. Figure S1 thus shows that the contact resonance is clearly visible in the QPDI measurement at x/L ≈ 0.6 (purple), whereas it is well below the noise floor for the OBD measurement at x/L ≈ 1 (red). As shown in the main text, this noise-floor improvement allows the QPDI detector to directly track the resonant frequency of the cantilever without an external drive.